\documentstyle[12pt, twocolumn]{article}

\newcommand{\be}{\begin{equation}}
\newcommand{\ee}{\end{equation}}

\newcommand{\bary}{\begin{eqnarray}}
\newcommand{\eary}{\end{eqnarray}}
\begin{document}
\title{Comment on ''Cerenkov radiation by neutrinos in a supernova core"} 
{\author{Subhendra Mohanty$^1$ and Sarira Sahu$^{2}$\\          
Theory Group, Physical Research Laboratory,\\
Navrangpura, Ahmedabad-380 009\\ India}
\date{ }
\maketitle
\footnotetext[1]{email:mohanty@prl.ernet.in}
\footnotetext[2]{email:sarira@prl.ernet.in}
\thispagestyle{empty}

It had been pointed out by Mohanty and Samal\cite{mohanty} 
that the
helicity flipping Cerenkov process $\nu_L\rightarrow\nu_R +\gamma$ or
$\gamma +\nu_L\rightarrow\nu_R$ could be an important cooling mechanism
for the supernova core. 
Comparing the neutrino emissivity by the Cerenkov process 
with observations of SN1987A, a restrictive bound  
on the neutrino magnetic moment was established. 
Subsequently it was pointed out by Raffelt\cite{raffelt} that,
this result was based on a numerical error in the calculation
of the refractive index of the SN core and using the correct numbers it was
shown that the photons in a SN core do not have a space-like 
dispersion relation, so the
Cerenkov helicity flip process would not occur. 

Here we show that the
earlier estimate of refractive index  was based on the thermodynamic formula
for susceptibility which turns out to be 
invalid for real photons or plasmons even in the
static limit. However an analysis of the dispersion relations of plasmons in
an ultrarelativistic plasma\cite{bratten} 
shows that the longitudinal photon (plasmon)
has a space-like branch and therefore in such a plasma the Cerenkov 
radiation of a plasmon is kinematically allowed\cite{sarira}. 
We compute the neutrino
emissivity of SN core by the Cerenkov helicity flip emission of a plasmon 
and subsequent escape of the $\nu_R$.
We show that the observations of neutrino flux from SN1987A put a constraint
on the neutrino magnetic moment $\mu_{\nu} < 0.6\times 10^{-11}\mu_B$.

The dispersion relation of a longitudinal photon (plasmon) is given by
\cite{bratten}
\be
k^2-\frac{3}{2}m^2_{\gamma}
\left [ \frac{w}{k}\ln \left(|\frac{w+k}{w-k}|\right )-2\right ]
=0.
\label{disp}
\ee
Using this dispersion relation we find that in a supernova core there is a
range of frequencies $(0.3 - 50)~ MeV$ for which the refractive index
$n=k/w > 1$ and therefore plasmon emission or absorbtion by the Cerenkov
process is kinematically allowed in this range .

The matrix element for the 
helicity flipping Cerenkov process $\nu_L\rightarrow\nu_R +\gamma$ is
\be
{\cal M} = \mu_{\nu} {\bar u}(p_2)\sigma^{\mu\nu}u(p_1)
\sqrt{Z_l}\epsilon_{\mu}(k, \lambda).
\ee
The photon polarisation sum in the medium for the longitudinal part is
given by
\be
\sum_{\lambda} \epsilon^L_{\mu}\epsilon^{L*}_{\nu}
=\left (1-\frac{1}{n^2}\right )u_{\mu}u_{\nu} +
\frac{(u_{\mu}k_{\nu} + k_{\mu}u_{\nu})}{n^2 w},
\ee
where $u_{\mu}$ is the four velocity of the medium.
The rate of energy emission by this process is
\be
{\dot S}=\frac{\mu^2_{\nu}}{16\pi E^2_1}
\int (E_1-w) w^2 dw Z_l f(n) 
(w-2 E_1)^2
\ee
where $f(n)=(n^2-1)^2/n^2$,  
the wave function renormalisation factor 
\be
Z_l=\frac{2}{n^2 + \frac{3 m^2_{\gamma}}{w^2}-1},
\ee
and $m_{\gamma}=e^2{\tilde\mu}^2/{3 \pi^2}$ is the photon mass in the
medium.
The neutrino luminosity is given by
\be
Q_{\nu_R}=V\int_0^{\infty} E^2_1 dE_1~ f_{\nu}(E_1) {\dot S}(E_1),
\ee
where $f_{\nu}(E_1)$ is the distribution function of the neutrino.
\begin{figure}
\vskip 6cm
\includegraphics{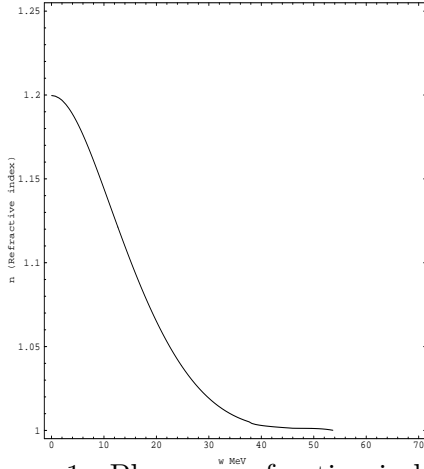}
\caption[dummy]{Plasmon refractive index in a supernova core as a function
of frequency}
\end{figure}

Taking the SN core volume to be
$V=4.19\times 10^{18}~ cm^3$, the electron chemical potential
$\tilde \mu_e=280$ MeV and the neutrino chemical potential 
$\tilde\mu_{\nu}=160$ MeV\cite{mohapatra} 
we obtain the neutrino luminosity to be
\be
Q_{\nu_R}=1.16\times 10^{60}\mu^2_{\nu}~MeV^4,
\ee
in terms of the magnetic moment of the neutrino. Assuming that the entire
energy of the core collapse is not carried away by the right handed 
neutrinos $i.e.$, $Q_{\nu_R} < 10^{52}~ergs/sec$ we obtain the upper 
bound on the neutrino magnetic moment 
$\mu_{\nu} <0.6\times10^{-11}\mu_B$.
This is comparable to the bound $(0.2 - 0.8)\times 10^{-11}\mu_B$
obtained\cite{mohapatra} from the cooling of SN1987A by the helicity flip
scattering.

\bibliographystyle{plain}

\end{document}